\documentclass[final,5p,times,twocolumn]{elsarticle}

\biboptions{sort&compress}

\usepackage{graphicx}
\usepackage{amssymb}
\usepackage{amsmath}
\usepackage{amsfonts}

\journal{Physics Letters A}

\begin{document}

\begin{frontmatter}

\title{An exact solution of the moving boundary problem for the expansion of a plasma cylinder in a
magnetic field}

\author{Hrachya B. Nersisyan}
\ead{hrachya@irphe.am}
\address{Division of Theoretical Physics, Institute of Radiophysics and Electronics, 0203
Ashtarak, Armenia}

\begin{abstract}
An exact analytic solution has been obtained for a uniformly expanding, neutral, infinitely conducting plasma
cylinder in an external uniform and constant magnetic field. The electrodynamical aspects related to the emission
and transformation of energy have been considered as well. The results obtained can be used in analysing the
recent experimental and simulation data.
\end{abstract}


\end{frontmatter}

\section{Introduction}
\label{sec:1}

Many processes in physics involve boundary surfaces which requires the solution of boundary and initial value problems.
The introduction of a moving boundary into the physics usually precludes the achievement of an exact analytic solution
of the problem and recourse to approximation methods is required \cite{rog89,mor53,jac75} (see also references
therein). In the case of a moving plane boundary a time-dependent translation of the embedding space immobilizes the
boundary at the expense of the increased complexity of the differential equation. It is the aim of this work to present
an example of a soluble moving boundary and initial value problem in cylindrical geometry.

The problems with moving boundary arise in many area of physics. One important example is sudden expansion of hot plasma
with a sharp boundary in an external magnetic field which is particularly of interest for many astrophysical and laboratory
applications (see, e.g., \cite{zak03} and references therein). Such kind of processes arise during the dynamics of solar
flares and flow of the solar wind around the earth's magnetosphere, in active experiments with plasma clouds in space,
and in the course of interpreting a number of astrophysical observations \cite{jac75,zak03,pon89,sag66,ler83,dit00,osi03,ner09,win05}.

To study the radial dynamics and evolution of the initially spherical or cylindrical plasma cloud both analytical and
numerical approaches were developed (see, e.g., Refs.~\cite{jac75,zak03,pon89,sag66,ler83,dit00,osi03,ner09,win05,kat61,rai63,ner06}
and references therein). The plasma cloud is shielded from the penetration of the external magnetic field by means of surface currents
circulating inside the thin layer on the plasma boundary. Ponderomotive forces resulting from interaction of these
currents with the magnetic field would act on the plasma surface as if there were magnetic pressure applied from
outside. After some period of accelerated motion, plasma gets decelerated as a result of this external magnetic
pressure acting inward. The plasma has been considered as a highly conducting media with zero magnetic field inside.
From the point of view of electrodynamics it is similar to the expansion of a superconductor in a magnetic
field. An exact analytic solution for a uniformly expanding, highly conducting plasma sphere in an external uniform
and constant magnetic field has been obtained in \cite{kat61}. The non-relativistic limit of this theory has been
used by Raizer \cite{rai63} to analyse the energy balance (energy emission and transformation) during the plasma
expansion. The similar problem has been considered in Ref.~\cite{dit00} within one-dimensional geometry for a plasma
layer. In our previous paper \cite{ner06} we obtained an exact analytic solution for the uniform relativistic expansion
of the highly conducting plasma sphere in the presence of a dipole magnetic field. In the present paper we study the
uniform expansion of the highly conducting plasma cylinder in the presence of a constant magnetic field. For this
geometry we found again an exact analytical solution which can be used in analysing the recent experimental and
simulation data (see, e.g., Refs.~\cite{zak03,win05} and references therein).

\section{Moving boundary problem}
\label{sec:2}

We consider the moving boundary problem of the highly conducting plasma cylinder expansion in the vacuum. Consider a
cylindrical region of space with radius $\rho =R(t)$ at the time $t$ containing a neutral infinitely conducting plasma
which has expanded at $t=0$ (with $R(0)=0$) to its present state from a linear source located at $\rho =0$. We assume
that at any time $t$ the plasma cylinder is unbounded in $z$ direction (i.e. the cylinder is located at $-\infty <z<\infty$).
To solve the boundary problem we introduce the cylindrical coordinate system $(\rho ,\varphi ,z)$ with the $z$-axis
along the plasma cylinder symmetry axis and the azimuthal angle $\varphi$ is counted from the plane ($xz$-plane)
containing the vector of the constant and homogeneous magnetic field $\mathbf{H}_{0}$. The angle $\vartheta$ between
the vector $\mathbf{H}_{0}$ and the $z$-axis is arbitrary. The unperturbed magnetic field is expressed by the vector
potential, $\mathbf{H}_{0}=\mathbf{\nabla }\times \mathbf{A}_{0}$, where this potential is $\mathbf{A}_{0}=\frac{1}{2}%
[\mathbf{H}_{0}\times \mathbf{r}]$ with the cylindrical components
\begin{eqnarray}
&&A_{0\rho} =-\frac{z}{2}H_{0\bot}\sin \varphi , \quad
A_{0z}=\frac{\rho }{2}H_{0\bot}\sin \varphi ,  \\
&&A_{0\varphi }=\frac{1}{2}\left(\rho H_{0\parallel}-zH_{0\bot}\cos \varphi \right) .  \nonumber
\label{20}
\end{eqnarray}
The components of the vector $\mathbf{H}_{0}$ are $H_{0\rho} =H_{0\bot}\cos \varphi$, $H_{0\varphi}=-H_{0\bot}\sin\varphi$,
and $H_{0z}=H_{0\parallel}$. Here $H_{0\bot}=H_{0}\sin \vartheta$ and $H_{0\parallel}=H_{0}\cos \vartheta$ are the components
of the magnetic field transverse and parallel to the $z$-axis, respectively.

As the cylindrical plasma cloud expands it both perturbs the external magnetic field and generates an electric field. Within the cylindrical
plasma region there is neither an electric field nor a magnetic field. We shall obtain an analytic solution of the electromagnetic field
configuration.

We consider the case of the uniform expansion of the plasma cylinder $R(t)=vt$ with a constant expansion velocity $v$. This special case
of the uniform expansion falls within the conical flow techniques which has been applied previously in Ref.~\cite{kat61}. From symmetry
considerations one seeks a solution for the total (i.e., the unperturbed potential $\mathbf{A}_{0}$ plus the induced one) vector potential
of the form $\mathbf{A}(\mathbf{r},t) \sim \rho^{\nu}\mathbf{a} (\zeta )$ with $\zeta =\rho /ct$, where $c$ is the velocity of light.
Since the external region of the conducting cylinder is devoid of free charge density, a suitable gauge allows the electric and magnetic
fields to be derived from the vector potential $\mathbf{A}$. Having in mind the symmetry of the unperturbed magnetic field and the fact
that the electromagnetic fields do not depend on the coordinate $z$ it is sufficient to choose the vector potential in the form
$A_{\rho }\left( \mathbf{r},t\right) =0$,
\begin{equation}
A_{\varphi }\left( \mathbf{r}%
,t\right) =\frac{H_{0\parallel}}{2}\rho \chi \left( \zeta \right) ,  \quad
A_{z}\left( \mathbf{r},t\right) =H_{0\bot }\rho \chi \left( \zeta \right)
\sin \varphi ,
\label{22}
\end{equation}%
where $\chi (\zeta)$ is some unknown function. The components of the electromagnetic field are expressed by this function as
\begin{eqnarray}
&&H_{\rho } =H_{0\bot }\chi \left( \zeta \right) \cos \varphi , \,\,
H_{\varphi } =-H_{0\bot }\sin \varphi \left[\zeta \chi \left(\zeta \right) \right]^{\prime} ,  \label{23a}    \\
&&H_{z} =H_{0\parallel}\left[\frac{\zeta }{2}\chi ^{\prime }\left( \zeta \right)
+\chi \left( \zeta \right) \right] ,    \nonumber  \\
&&E_{\varphi }=\frac{H_{0\parallel}}{2}\zeta ^{2}\chi ^{\prime }\left( \zeta \right) , \,\,
E_{z}=H_{0\bot }\zeta ^{2}\chi^{\prime }\left( \zeta \right) \sin \varphi ,  \label{23}
\end{eqnarray}%
$E_{\rho}=0$, where the prime indicates the derivative with respect to the argument. The equation for the vector potential $\mathbf{A}
(\mathrm{r},t)$ is obtained from the Maxwell's equations which for the unknown function $\chi (\zeta)$ yields an ordinary differential
equation
\begin{equation}
\chi ^{\prime \prime }\left( \zeta \right) +\frac{3-2\zeta ^{2}}{\zeta
\left( 1-\zeta ^{2}\right) }\chi ^{\prime }\left( \zeta \right) =0 .
\label{24}
\end{equation}
This equation is to be solved in the external region $\rho >R(t)$ subject to the boundary and initial conditions.
Here $R(t)$ is the plasma cylinder radius at the time $t$. The initial conditions are at $t=0$
\begin{equation}
\mathbf{A}\left(\mathbf{r},0\right) =\mathbf{A}_{0} ,  \quad
\frac{\partial \mathbf{A}\left(\mathbf{r},0\right)}{\partial t}=0 .
\label{25}
\end{equation}%
The first initial condition states that the initial value of $\mathbf{A}(\mathrm{r},t)$ is that of a homogeneous
unperturbed magnetic field. The second initial condition states that there is no initial electric field. Boundary
conditions should be imposed at the cylindrical surface $\rho =R(t)$ and at infinity. Because of the finite
propagation velocity of the perturbed electromagnetic field the magnetic field at infinity will remain undisturbed
for all finite times. Further, no incoming wave-type solutions are permitted. Thus, for all finite times
$\mathbf{H}(\mathbf{r},t) \to \mathbf{H}_{0}$ at $\rho\to \infty$. The boundary condition at the expanding cylindrical
surface is $H_{\rho}=0$ which is equivalent to the relation $\chi (\beta ) =0$ (see Eq.~\eqref{23a}) with $\beta =v/c$. In addition imposing
that $\mathbf{H}(\mathbf{r},t)=\mathbf{H}_{0}$ at $\rho\geqslant ct$ we obtain another boundary condition $\chi (1)=1$.

The solution of Eq.~\eqref{24} subject to the initial and boundary conditions may be written as
\begin{equation}
\chi \left(\zeta \right) =1-\frac{\mathcal{F}\left(\zeta \right)}{\mathcal{F}\left(\beta\right)} ,
\label{30}
\end{equation}%
where
\begin{equation}
\left(
\begin{array}{c}
\mathcal{U}(\zeta ) \\
\mathcal{F}(\zeta )%
\end{array}%
\right) =\frac{\sqrt{1-\zeta ^{2}}}{\zeta ^{2}} \pm \ln \frac{1+ \sqrt{1-\zeta ^{2}}}{\zeta } .
\label{31}
\end{equation}%

The complete solution may finally be written in the form at $vt\leqslant \rho\leqslant ct$ (or $\beta\leqslant \zeta
\leqslant 1$)
\begin{eqnarray}
H_{\rho} =H_{0\perp}\cos \varphi \left[ 1-\frac{\mathcal{F}\left(\zeta \right)}{\mathcal{F}\left(\beta \right)}\right] ,  \nonumber \\
H_{\varphi} =-H_{0\perp}\sin \varphi \left[1+\frac{\mathcal{U}\left(\zeta\right)}{\mathcal{F}\left(\beta \right)}\right] ,  \label{34a} \\
H_{z} =H_{0\parallel}\left[1+\frac{1}{\mathcal{F}\left( \beta \right) }\ln \frac{1+\sqrt{1-\zeta ^{2}}}{\zeta }\right] ,   \nonumber \\
E_{\rho}=0 ,  \quad  E_{\varphi } =\frac{H_{0\parallel}}{\mathcal{F}\left(\beta \right)}\frac{\sqrt{1-\zeta ^{2}}}{\zeta } ,  \label{34}  \\
E_{z} =H_{0\perp}\sin \varphi \frac{2\sqrt{1-\zeta ^{2}}}{\mathcal{F}\left(\beta\right) \zeta} ,  \nonumber
\end{eqnarray}%
$\mathbf{H}(\mathbf{r},t) =\mathbf{H}_{0}$, $\mathbf{E}(\mathbf{r},t) =0$ at $\rho >ct$ (or $\zeta >1$) and
$\mathbf{H}(\mathbf{r},t) =\mathbf{E}(\mathbf{r},t) =0$ at $\rho <R(t)=vt$ (or $\zeta <\beta$).  From Eqs.~\eqref{23a},
\eqref{23}, \eqref{34a} and \eqref{34} it can be easily checked that the boundary condition on the moving surface,
$\mathbf{E}(R) =-\frac{1}{c}[\mathbf{v} \times \mathbf{H}(R) ]$ (or $E_{\varphi}(R) =\beta H_{z}(R)$, $E_{z}(R) =
-\beta H_{\varphi}(R)$), is satisfied automatically.

It is also imperative to determine the surface current density induced due to the moving boundary. This can be
done employing the Maxwell's equation $\mathbf{j} =\frac{1}{4\pi}\left(c\nabla\times \mathbf{H}-\frac{\partial
\mathbf{E}}{\partial t}\right)$. It is clear that the surface current has only two components and $j_{\alpha}=i_{\alpha}\delta
(\rho -R)$ with $\alpha =z, \varphi$, where $i_{\alpha}$ is the linear surface current. Using Eqs.~\eqref{34a}
and \eqref{34} as well as the Maxwell's equation we obtain
\begin{eqnarray}
&&i_{z} =\frac{c}{4\pi }\left[ H_{\varphi }\left( R\right)
+\beta E_{z}\left( R\right) \right] =\frac{c}{4\pi \gamma ^{2}}H_{\varphi}\left( R\right)  \nonumber \\
&&=-i_{0\perp }\sin \varphi \frac{2}{\gamma ^{3}\beta ^{2}%
\mathcal{F}\left( \beta \right)} ,  \label{eq:curr}   \\
&&i_{\varphi } =-\frac{c}{4\pi }\left[ H_{z}\left( R\right)
-\beta E_{\varphi }\left( R\right) \right] =-\frac{c}{4\pi \gamma ^{2}} H_{z}\left( R\right)  \nonumber  \\
&&=-i_{0\parallel }\frac{1}{\gamma ^{3}\beta ^{2}\mathcal{F}\left( \beta \right)} . \label{eq:curr1}
\end{eqnarray}
Here $\gamma^{-2}=1-\beta^{2}$ is the relativistic factor of the expanding boundary and $i_{0\parallel ,\perp }=
(c/4\pi )H_{0\parallel ,\perp }$. Note that the moving boundary modifies the surface current which now contains
an extra factor $\gamma^{-2}$ \cite{jac75}.

Consider now briefly the non-relativistic limit of Eqs.~\eqref{34a} and \eqref{34}. This limit can be obtained using at
$\zeta \to 0$ and $\beta \to 0$ the asymptotic expression $\mathcal{U}(\zeta )/\mathcal{F}(\beta )=\mathcal{F}(\zeta )%
/\mathcal{F}(\beta ) =\beta^{2}/\zeta^{2}=R^{2}/\rho^{2}$ which yields $H_{z} =H_{0\parallel}$, and
\begin{equation}
H_{\rho} =H_{0\perp}\cos \varphi \left( 1-\frac{R^{2}}{\rho^{2}}\right) ,  \,\,
H_{\varphi} =-H_{0\perp}\sin \varphi \left(1+\frac{R^{2}}{\rho^{2}}\right) .
\label{eq:nr}
\end{equation}
In the lowest order with respect to the factor $\beta$ the components of the electric field are given by $E_{\varphi} =%
\beta H_{0\parallel} (R/\rho )$, $E_{z} =2\beta H_{0\perp}\sin \varphi (R/\rho )$. It is seen that the parallel
component $H_{z}$ of the magnetic field remains unchanged in the case of non-relativistic expansion.

\section{Analysis of the energy balance}
\label{sec:3}

Previously significant attention has been paid \cite{dit00,osi03,ner09,rai63,ner06} to the question of what fraction of energy
is emitted and lost in the form of electromagnetic pulse propagating outward of the expanding plasma. In this
section we consider the energy balance during the plasma cylinder expansion in the presence of the homogeneous
magnetic field. When the plasma cylinder of the zero initial radius is created at $t=0$ and starts expanding,
external magnetic field $\mathbf{H}_{0}$ is perturbed by the electromagnetic pulse, $\mathbf{H}^{\prime }(\mathbf{r},t)
=\mathbf{H}(\mathbf{r},t)-\mathbf{H}_{0}$, $\mathbf{E}(\mathbf{r},t)$, propagating outward with the speed of
light. The tail of this pulse coincides with the moving plasma boundary $\rho =R(t)$ while the leading edge is at
$\rho =ct$. Ahead of the leading edge, the magnetic field is not perturbed and equals $\mathbf{H}_{0}$ while the
electric field is zero.

Our starting point is the energy balance equation (Poynting equation)
\begin{equation}
\mathbf{\nabla }\cdot \mathbf{S}=-\mathbf{j}\cdot \mathbf{E}-\frac{\partial
}{\partial t}\frac{E^{2}+H^{2}}{8\pi } ,
\label{35}
\end{equation}%
where $\mathbf{S}=\frac{c}{4\pi }[\mathbf{E}\times \mathbf{H}]$ is the Poynting vector and $\mathbf{j}=j_{\varphi }%
\mathbf{e}_{\varphi }+j_{z}\mathbf{e}_{z}$ (with $\vert \mathbf{e}_{\varphi }\vert =\vert \mathbf{e}_{z}\vert=1$) is
the surface current density. The energy emitted to infinity is measured as a Poynting vector integrated over time
and over the lateral surface $S_{c}$ of the cylinder with radius $\rho_{c}$, length $l_{c}$ and the volume $\Omega_{c}$
(control cylinder) enclosing the plasma cylinder ($\rho_{c}>R$ or $0\leqslant t<\rho_{c}/v$). Integrating over time
and over the volume $\Omega_{c}$ Eq.~\eqref{35} can be represented as
\begin{equation}
W_{S}\left( t\right) =W_{J}\left( t\right) +\Delta W_{\mathrm{EM}}\left(t\right) ,
\label{36}
\end{equation}%
where
\begin{equation}
 W_{S}\left( t\right) =l_{c}\rho_{c} \int_{0}^{t}dt^{\prime}\int_{0}^{2\pi }S_{\rho} d\varphi ,  \,\,
 W_{J}\left( t\right)=-\int_{0}^{t}dt^{\prime }\int_{\Omega _{c}}\mathbf{j}\cdot \mathbf{E}d\mathbf{r} .
\label{37}
\end{equation}%
Here $S_{\rho}=\frac{c}{4\pi }(E_{\varphi }H_{z}-E_{z} H_{\varphi})$ is the radial component of the Poynting vector.
Note that the total flux of the energy over the bases of the control cylinder determined by the Poynting's vector
component $S_{z}$ vanishes due to the symmetry reason.
$W_{\mathrm{EM}}(t)$ and $\Delta W_{\mathrm{EM}}(t)=W_{\mathrm{EM}}(0) -W_{\mathrm{EM}}(t)$ are the total electromagnetic
energy and its change (with minus sign) in a volume $\Omega _{c}$, respectively. $W_{J}(t)$ is the energy transferred
from plasma cylinder to electromagnetic field and is the mechanical work with minus sign performed by the plasma on the
external electromagnetic pressure. At $t=0$ the electromagnetic fields are given by $\mathbf{H}(\mathbf{r},t) =\mathbf{H}_{0}$
and $\mathbf{E}(\mathbf{r},t)=0$. Hence $W_{\mathrm{EM}}(0)$ is the total energy of the magnetic field in a volume $\Omega _{c}$
and is given by $W_{\mathrm{EM}}(0) \equiv Q =\pi \rho^{2}_{c} l_{c} (H_{0}^{2}/8\pi )$. Then the change of the electromagnetic
energy $\Delta W_{\mathrm{EM}}(t)$ in a volume $\Omega _{c}$ can be evaluated as
\begin{equation}
\Delta W_{\mathrm{EM}}\left( t\right) =Q-\int_{\Omega_{c}^{\prime }}\frac{E^{2}+H^{2}}{8\pi }d\mathbf{r} .
\label{39}
\end{equation}%
In Eq.~\eqref{39} $\Omega_{c}^{\prime }$ is the volume of the control cylinder excluding the volume of the plasma
cylinder (we take into account that $\mathbf{H}(\mathbf{r},t)=\mathbf{E}(\mathbf{r},t)=0 $ in a plasma cylinder). Hence
the total energy flux $W_{S}(t)$ given by Eq.~\eqref{37} is calculated as a sum of the energy loss by plasma due
to the external electromagnetic pressure and the decrease of the electromagnetic energy in a control volume $\Omega_{c}$.
For non-relativistic ($\beta \ll 1$) expansion of a one-dimensional plasma slab and for uniform external magnetic field
$W_{S}\simeq 2W_{J} \simeq 2\Delta W_{\rm EM}$, i.e., approximately the half of the outgoing energy is gained from the
plasma, while the other half is gained from the magnetic energy \cite{dit00}. In the case of non-relativistic expansion
of highly-conducting spherical plasma with radius $R$ in the uniform magnetic field $\mathbf{H}_{0}$ the outgoing energy
$W_{S}$ is distributed between $W_{J}$ and $\Delta W_{\mathrm{EM}}$ according to $W_{J}=1.5Q_{0}$ and $\Delta W_{\mathrm{EM}}%
=0.5Q_{0}$ with $W_{S}=2Q_{0}$, where $Q_{0}=H_{0}^{2}R^{3}/6$ is the magnetic energy escaped from the spherical plasma
volume \cite{rai63}. Therefore in this case the released electromagnetic energy is mainly gained from the plasma.

Consider now each energy component $W_{S}(t)$, $W_{J}(t)$ and $\Delta W_{\mathrm{EM}}(t)$ separately. $W_{S}(t)$ is
calculated from Eq.~\eqref{37}. In the first expression of Eq.~\eqref{37} the $t^{\prime }$-integral must be performed
at $\frac{\rho_{c}}{c}\leqslant t^{\prime }\leqslant t$ ($t<\frac{\rho_{c}}{v}$) since at $0\leqslant t^{\prime}<\frac%
{\rho_{c}}{c}$ the electromagnetic pulse does not reach to the control surface yet and $S_{\rho}(\rho_{c})=0$. From
Eqs.~\eqref{34a}, \eqref{34} and \eqref{37} we obtain
\begin{eqnarray}
&&\frac{W_{S}\left( t\right) }{Q}=\frac{1}{\mathcal{F}^{2}\left( \beta \right) }\Bigg[
2\mathcal{F}\left(\beta \right) \mathcal{F}\left(\eta \right) -\mathcal{F}^{2}\left(\eta \right)  \nonumber \\
&&+\left( 1+\frac{H_{0\perp }^{2}}{H_{0}^{2}}\right) \frac{\left( 1-\eta
^{2}\right) ^{2}}{\eta ^{4}}\Bigg] , \label{40}
\end{eqnarray}
where $\eta =\rho_{c}/ct<1$. In non-relativistic limit using the asymptotic expression $\mathcal{F}(\eta)/%
\mathcal{F}(\beta)=\beta^{2}/\eta^{2} =(t/t_{v})^{2}$ (with $t_{v} =\rho_{c}/v$) at $\beta \to 0$ and $\eta\to 0$,
from Eq.~\eqref{40} we obtain
\begin{equation}
\frac{W_{S}\left( t\right) }{Q}=\frac{2t^{2}}{t^{2}_{v}} +\frac{t^{4}}{t^{4}_{v}}\frac{H_{0\perp}^{2}}{H_{0}^{2}} .
\label{42}
\end{equation}%

\begin{figure*}[tbp]
\begin{center}
\includegraphics[width=130.0mm]{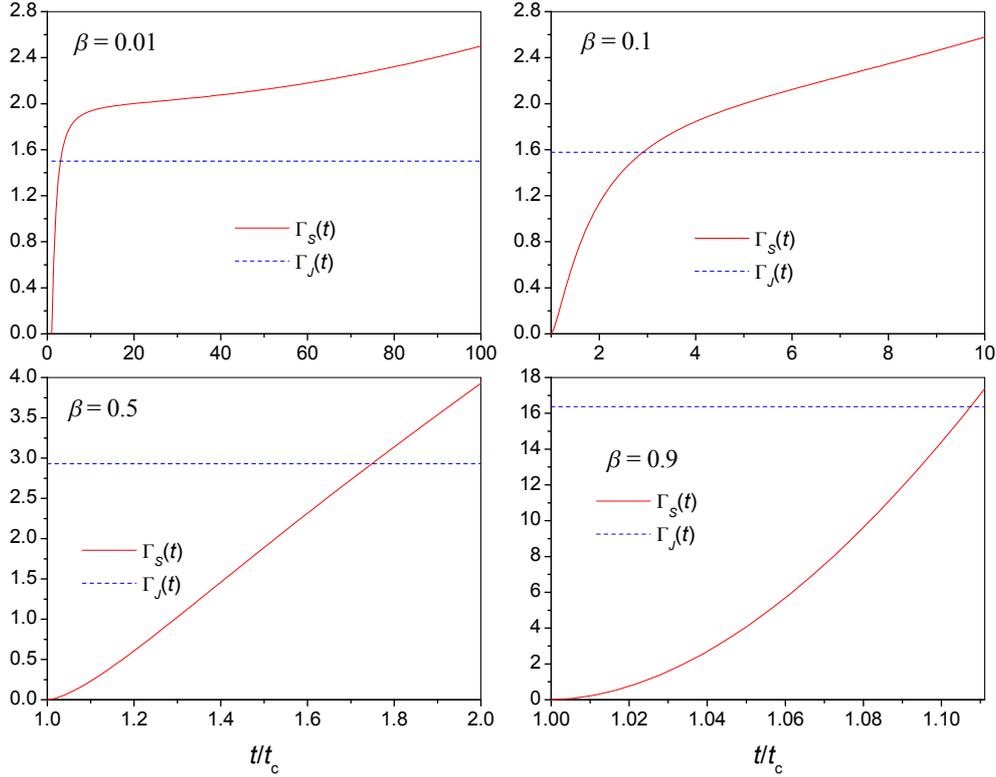}
\end{center}
\caption{(Color online.) The ratios $\Gamma _{S}(t)$ (solid lines) and $\Gamma _{J}(t)$ (dashed lines) for four values of $\beta $ as
a function of $t$ (in units of $t_{c}=\rho_{c}/c$) calculated from expressions \eqref{40} and \eqref{44} with $\vartheta =\pi /4$.}
\label{fig:energy}
\end{figure*}

Next, we evaluate the energy loss $W_{J}(t)$ by the plasma which is determined by the surface current density, $\mathbf{j}$.
This current has two azimuthal and axial components and is localized within thin cylindrical skin layer, $R-\delta <\rho <R+\delta$
with $\delta \to 0$, near plasma boundary. Therefore in Eq.~\eqref{37} the volume $\Omega_{c}$ can be replaced by the
volume $\Omega_{\delta }\sim l_{c}R\delta $ which includes the space between the cylinders with $\rho =R-\delta$ and $\rho =R+\delta$.
The surface current density is calculated from the Maxwell's equation and has been determined in previous section, see Eqs.~\eqref{eq:curr}
and \eqref{eq:curr1}. As shown below the $\mathbf{r}$-integration of the term $\mathbf{j}\cdot \mathbf{E}$ in Eq.~\eqref{37} can be
alternatively expressed via magnetic field. Within the skin layer we take into account that $\mathbf{E}=-\frac{1}{c}
[\mathbf{v}\times \mathbf{H}]$ and $H_{\rho}(R)=0$. Then
\begin{eqnarray}
&&Q_{J}\left( t\right) =-\int_{\Omega _{\delta }}\mathbf{j}\cdot \mathbf{%
E}d\mathbf{r}=\frac{1}{4\pi }\int_{\Omega _{\delta }}\mathbf{v\cdot }\left[
\mathbf{H}\times \left( \mathbf{\nabla }\times \mathbf{H}\right) \right] d\mathbf{r}  \nonumber \\
&&+\frac{1}{8\pi }\int_{\Omega _{\delta }}\frac{\partial \mathbf{E}%
^{2}}{\partial t}d\mathbf{r} =v\int\nolimits_{S_{R}}\frac{H^{2}\left( R\right) -E^{2}\left( R\right) }{8\pi }dS  \label{43}  \\
&&=\frac{v}{\gamma ^{2}}\int\nolimits_{S_{R}}\frac{H^{2}\left( R\right) }{8\pi }dS,    \nonumber
\end{eqnarray}%
where $S_{R}$ is the lateral surface of the expanding cylindrical plasma. As mentioned in Sec.~\ref{sec:2} the factor $\gamma^{-2}$
in Eq.~\eqref{43} appears because of the moving boundary. In this expression
the term with $\frac{\partial \mathbf{E}^{2}(\mathbf{r},t)}{\partial t}$ has been transformed to the surface integral using the
fact that the boundary of the volume $\Omega_{\delta }$ moves with a constant velocity $v$ and the electrical field has a jump
across the plasma surface. Equation~\eqref{43} shows that the energy loss by the plasma per unit time is equal to the work
performed by the plasma on the external electromagnetic pressure. This external pressure is formed by the difference between
magnetic and electric pressures, i.e., the induced electric field tends to decrease the force acting on the expanding plasma
surface. The total energy loss by the plasma cylinder is calculated as
\begin{equation}
\frac{W_{J}\left( t\right) }{Q}=\frac{1}{Q}\int_{0}^{t}Q_{J}\left( t^{\prime
}\right) dt^{\prime }=\left( 1+\frac{H_{0\perp }^{2}}{H_{0}^{2}}\right)
\frac{\left(1-\beta^{2}\right)^{2}}{\eta^{2} \beta ^{2} \mathcal{F}^{2}\left( \beta \right)} .
\label{44}
\end{equation}%
In a non-relativistic case Eq.~\eqref{44} yields:
\begin{equation}
\frac{W_{J}\left( t\right)}{Q}=\frac{t^{2}}{t^{2}_{v}} \left( 1+\frac{H_{0\perp }^{2}}{H_{0}^{2}}\right) .
\label{45}
\end{equation}

The change of the electromagnetic energy in a control cylinder is calculated from Eq.~\eqref{39}. At $R<\rho_{c} <ct$
(the electromagnetic pulse fills the whole control cylinder) we obtain
\begin{eqnarray}
&&\frac{\Delta W_{\mathrm{EM}}\left( t\right) }{Q}=\frac{1}{\mathcal{F}^{2}\left( \beta
\right) }\Bigg[2\mathcal{F} \left(\beta \right) \mathcal{F}\left(\eta \right) -\mathcal{F}^{2}\left(\eta \right)  \label{46}  \\
&&-\left( 1+\frac{H_{0\perp }^{2}}{H_{0}^{2}}\right) \frac{\eta ^{2}-\beta ^{2}}{\eta ^{2}}\left( \frac{1}%
{\beta ^{2}\eta ^{2}}-1\right) \Bigg] .  \nonumber
\end{eqnarray}%
Comparing Eqs.~\eqref{40}, \eqref{44} and \eqref{46} we conclude that $\Delta W_{\mathrm{EM}}(t)+W_{J}(t) =W_{S}(t)$ as
predicted by the energy balance equation~\eqref{36}. The non-relativistic limit of Eq.~\eqref{46} can be evaluated from
Eqs.~\eqref{42} and \eqref{45} using the relation $\Delta W_{\mathrm{EM}}(t) =W_{S}(t) -W_{J}(t)$. As an example in
Fig.~\ref{fig:energy} we show the results of model calculations for the ratios $\Gamma_{S}(t)=W_{S}(t)/Q_{0}(t)$ and $\Gamma_{J}(t)
=W_{J}(t)/Q_{0}(t)$ as a function of time ($\rho_{c}/c\leqslant t\leqslant \rho_{c}/v$). Here $Q_{0}(t)=\pi l_{c}R^{2}H^{2}_{0}/8\pi$ is
the magnetic energy escaped from the plasma cylinder at time $t$. For the relativistic factor $\beta $ we have chosen a wide range
of values. We recall that at $0\leqslant t\leqslant \rho_{c}/c$, i.e. the electromagnetic pulse does not yet reach to
the surface of the control cylinder, $W_{S}(t)=0$. Unlike the case with uniform magnetic field discussed above (see also
\cite{dit00,rai63}) there are no simple relations between the energy components $W_{S}(t)$, $W_{J}(t)$ and $Q_{0}(t)$.
From Eq.~\eqref{44} it is seen that the ratio $\Gamma _{J}(t)$ is constant and is given by $\Gamma _{J}(t)=(1+H_{0\perp}%
^{2}/H_{0}^{2})C$, where $C=[\gamma^{2}\beta^{2} \mathcal{F}(\beta )]^{-2}$ is some kinematic factor. For non-relativistic
expansion with $\beta\ll 1$, this factor is $C\simeq 1$ while at $\beta \sim 1$ this factor is very large and behaves as
$C\simeq (9/8) (1-\beta )^{-1}\gg 1$. As expected the total energy flux, $W_{S}(t)$, increases monotonically with $t$.
At the final stage ($t=\rho_{c}/v$) of relativistic expansion (with $\beta \sim 1$) $W_{S}\simeq W_{J}$. Hence in this
case the emitted energy $W_{S}$ is mainly gained from the plasma cylinder.

\section{Conclusion}
\label{sec:4}

An exact solution of the uniform radial expansion of a neutral, infinitely conducting plasma cylinder in the presence of a
homogeneous magnetic field has been obtained. The electromagnetic fields are derived by using the appropriate boundary and
initial conditions, Eq.~\eqref{25}. It has been shown that the electromagnetic fields are perturbed only within
the domain extending from the surface of the expanding plasma cylinder $\rho =R=vt$ to the surface of the expanding information
cylinder $\rho =ct$. External to the cylinder $\rho =ct$ the magnetic field is not perturbed. In the course of this study
we have also considered the energy balance during the plasma cylinder expansion. The model calculations show that the emitted
energy is mainly gained from the plasma cylinder. For relativistic expansion $W_{S} \simeq W_{J}$ and the emitted energy is
practically gained only from plasma cylinder.

We expect our theoretical findings to be useful in experimental investigations as well as in numerical simulations of the plasma
expansion into ambient magnetic field. One of the improvements of our model will be to include the effect of the deceleration
of the plasma cylinder as well as the derivation of the dynamical equation for the surface deformation. A study of this and
other aspects will be reported elsewhere.

\section*{Acknowledgements}
This work has been supported by the Armenian National Science and Educational Foundation (ANSEF),
Project PS 2183.

\end{document}